\documentclass[journal]{IEEEtran}

\usepackage[]{algorithm2e}
\usepackage{tikz}
\graphicspath{{images/}}
\usepackage{amsmath}
\usepackage{latexsym, amssymb}
\usepackage{graphicx}
\usepackage{amsmath, amsbsy}
\usepackage{amsopn, amstext}
\usepackage{ifpdf,hyperref}
\usepackage{cancel, color}
\usepackage{epstopdf}

\def\supp{supp}

\DeclareMathOperator{\Col}{Col}
\DeclareMathOperator{\Row}{Row}

\DeclareMathOperator{\lcm}{lcm}

\def\cal{\mathcal}

\def\diag{diag}
\def\ra{\rightarrow}

\def\lra{\leftrightarrow}

\def\a{\alpha}

\def\d{\delta}

\def\D{\Delta}

\def\0{{\bf 0}}

\def\dsum{\mathop{\sum}\limits}

\newtheorem{thm}{Theorem}[section]
\newtheorem{dfn}[thm]{Definition}
\newtheorem{prp}[thm]{Proposition}
\newtheorem{exa}[thm]{Example}

\newtheorem{rem}[thm]{Remark}

\begin{document}

\title{Hidden Order of Boolean Networks}

\author{Xiao Zhang, Zhengping Ji, Daizhan Cheng,~\IEEEmembership{Fellow,~IEEE}
\thanks{X. Zhang is with the National Center for Mathematics and Interdisciplinary Sciences \& Key Laboratory of Systems and Control, Academy of Mathematics and Systems Science, Chinese Academy of Sciences, Beijing 100190, P.R.China, e-mail: xiaozhang@amss.ac.cn}
\thanks{Jhengping Ji is with the Key Laboratory of Systems and Control, Academy of Mathematics and Systems Science \& School of Mathematical Sciences, University of Chinese Academy of Sciences, Beijing 100190, P.R.China, e-mail: jizhengping@amss.ac.cn }
\thanks{D. Cheng is with the Academy of Mathematics and Systems Science, Chinese Academy of Sciences, Beijing 100190, P.R.China, e-mail: dcheng@iss.ac.cn}
	\thanks{This work is supported partly by the National Natural Science Foundation of China (NSFC) under Grants 62073315, 61074114, and 61273013.}

}

\maketitle

\begin{abstract}

It is a common belief that the order of a Boolean network is mainly determined by its attractors, including fixed points and cycles. Using semi-tensor product (STP) of matrices and the algebraic state-space representation (ASSR) of Boolean networks, this paper reveals that in addition to this explicit order, there is certain implicit or hidden order, which is determined by the fixed points and limit cycles of their dual networks. The structure and certain properties of dual networks are investigated. Instead of a trajectory, which describes the evolution of a state, hidden order provides a global picture to describe the evolution of the overall network. It is our conjecture that the order of networks is mainly determined by the dual attractors via their corresponding hidden orders. The previously obtained results about Boolean networks are further extended to the $k$-valued case.
\end{abstract}


\IEEEpeerreviewmaketitle

\section{Introduction}

Boolean network (BN) was firstly proposed by Kauffman to describe genetic regulatory networks \cite{kau69}. It has then been proved efficient and attracted considerable attention from biologists, computer and system scientists, etc.
Kauffman's eventual purpose in proposing BN is ``to answer the question, what are the sources of the overwhelming and beautiful order which graces the living world?" \cite{kau93}. Roughly speaking, the answer is:
Nature selection, as proposed by Darwin, plus the ``emergent order" or ``order for free" from self-organization that arises naturally.

How does self-organization emerge in a network? Kauffman's viewpoint is: ``Tiny attractors" lead to ``vast, vast order". It was proposed in \cite{wal95} that: ``Under the right conditions, these attractors can be the source of order in large dynamical systems. Since the system follows trajectories that inevitably flow into attractors, tiny attractors will `trap' the system into tiny subregions of its state space. Among the vast range of possible behaviors, the system settles into an orderly few. The attractors, if small, create order. Indeed, tiny attractors are a prerequisite for the order for free that we are seeking."

Roughly speaking, the attractors, including fixed points and limit cycles, form the fundamental topological structure of the network. Hence, they determine the ``order" of mostly large-scale networks, which are particularly from metabolic networks or genetic regularity networks, etc.

In his book ``Hidden Order" \cite{hol95}, Holland described how the hidden order emerges from complex systems via adaptation, and how the hidden order determines the behavior of a complex system, which was said to be ``complexity made simple".

In a large-scale BN, such as a genetic regulatory network, the topological structure of
a BN is mainly determined by its fixed points and cycles, which are called attractors. Since a BN  has only finite nodes which leads to finite states, a trajectory starting from any state will converge to an attractor. Searching the order in lives through its attractors such as what Kauffman did seems reasonable \cite{kau93}.

Recently, we found a kind of hidden order of BNs, which may more significantly characterize certain properties of a BN. The hidden orders come from the fixed points and cycles of logical functions under the action of the structure matrix of BNs. It has been discovered that this hidden order is very important for revealing certain properties of a large-scale BN by constructing a much smaller realization, which involves only much smaller set of related states \cite{chepr}.

When $k$-valued logical networks are considered, similarly to BN, the arguments about the hidden order of BNs are also efficient for $k$-valued networks.

The notations used in the text is shown in TABLE \ref{notation}.

\begin{table}[h]
\renewcommand{\arraystretch}{1.3}
\caption{Notations used in this paper.}
\label{notation}
\centering
\begin{tabular}{l l}
  \hline
  Notation & Description\\
  \hline
    ${\cal M}_{m\times n}$ & The set of $m\times n$ real matrices.\\
    $\Col(A)$ ($\Row(A)$) & The set of columns (rows) of matrix $A$.\\ 
    $\Col_i(A)$ ($\Row_i(A)$) & The $i$-th column (row) of matrix $A$.\\
    ${\cal D}_k$ & $\{0,\frac{1}{k-1},\frac{2}{k-1},\cdots,1\}, \quad k\geq 2$.\\
    ${\cal D}_2$ (or ${\cal D}$) & $\{0,1\}$.\\
    $\delta_n^i$ & The $i$-th column of identity matrix $I_n$.\\
    $\Delta_n$ & $\Delta_n=\{\d_n^i\,|\,i=1,2,\cdots,n\}$.\\
    $[A]_{i,j}$ & $\Row_i(\Col_j(A))$.\\
    ${\cal L}_{m\times n}$ & The set of ${m\times n}$ logical matrices, that\\~& $\Col({\cal L})\subset \Delta_{m}$.\\
    ${\cal B}_{m\times n}$ & The set of $m\times n$ Boolean matrices, that\\~& $[{\cal B}]_{i,j}\subset {\cal D}$.\\
    ${\bf 1}_{n}$ & ${\underbrace{[1,1,\cdots,1]}_n}^\mathrm{T}$.\\
  ${\bf 0}_{m\times n}$ (or ${\bf 0}$) & The $m\times n$ (or default) zero matrix.\\
  $A\otimes B$ & The Kronecker product of matrices\\~& $A\in{\cal M}_{m\times n}$ and $B\in{\cal M}_{p\times q}$.\\
  $A*B$ & The Khatri-Rao product of $A \in{\cal M}_{p\times n}$, \\~&$B\in{\cal M}_{q\times n}$, that for $i=1,2,\cdots n$, \\~&$\Col_i(A*B)=\Col_i(A)\otimes\Col_i(B)$.\\
  \hline
\end{tabular}
\end{table}

The rest of this paper is organized as follows: The state space and its dual space of a BN are clarified in Section II. Section III explores the hidden order of a BN through its dual network. The attractors and dual attractors of a BN are discussed in Section IV, using the canonical form of the BN. Section V considers the hidden order determined by dual attractors. Section VI considers the Boolean algebraic structure on the dual space. The realization of a Boolean control network (BCN) is discussed in Section VII. In Section VIII a brief discussion is given for extending the technique developed for BNs to $k$-valued logical networks. Section IX is some concluding remarks which shows why the hidden attractors determined by dual MNs might play a more important role in determine the order of BNs.

\section{Dual Space of BN}

Consider an $n$-node BN, whose logical evolutionary dynamics is
\begin{align}\label{2.1.1}
\begin{cases}
X_1(t+1)=f_1(X_1,X_2,\cdots,X_n)\\
X_2(t+1)=f_2(X_1,X_2,\cdots,X_n)\\
\quad\quad\vdots\\
X_n(t+1)=f_n(X_1,X_2,\cdots,X_n),\\
\end{cases}
\end{align}
where $X_i(t)\in {\cal D}=\{0,1\},~i\in[1,n]$ are the nodes, and $f_i:{\cal D}^n\ra {\cal D}$ are logical functions.

Using vector expressions $1\sim \d_2^1$ and $0\sim \d_2^2$, $X_i$ can be expressed into its vector form as
$$
\vec{X}_i:=x_i=\begin{bmatrix}
X_i\\1-X_i\end{bmatrix},\quad i\in[1,n].
$$
Denote by $X=(X_1,X_2,\cdots,X_n)\in {\cal D}^n$ the overall state variable. Its vector form expression is
$$
\vec{X}:=x=\ltimes_{i=1}^nx_i.
$$

Assume $M\in {\cal L}_{m\times n}$. By definition, $M=[\d_m^{i_1},\d_m^{i_2},\cdots,\d_m^{i_n}]$. Its condensed form is
\begin{align}\label{2.1.101}
M=\d_{m}[i_1,i_2,\cdots,i_n].
\end{align}

The mathematical tool used in this paper is the semi-tensor product of matrices which is defined as below.
\begin{dfn}\cite{che11}\label{d2.1.1} Let ~$A\in {\cal M}_{m\times n}$,  $B\in {\cal M}_{p\times q}$, and the least common multiple of $n$ and $p$ be $t=\lcm\{n,p\}$.
The semi-tensor product of $A$ and $B$, denoted by $A\ltimes B$, is defined as
\begin{equation} \left(A\otimes I_{t/n}\right)\left(B\otimes I_{t/p}\right).
\end{equation}
\end{dfn}

The following results are borrowed from \cite{che11,che12}.
\begin{prp}\label{p2.1.1}
\begin{itemize}
\item[(i)] Let $f:{\cal D}^n\ra {\cal D}$, expressed by $Y=f(X_1,X_2,\cdots,X_n)$, be a Boolean function. Then there exists a unique logical matrix $M_f\in {\cal L}_{2\times 2^n}$, named the structure matrix of $f$, such that in vector form ($y=\vec{Y}$) the Boolean function can be expressed by
\begin{align}\label{2.1.3}
y=M_fx.
\end{align}
\item[(ii)] Let $M_i$ be the structure matrix of Boolean functions $f_i$, $i=1,2,\cdots,n$. Then there exists a unique logical matrix $M\in {\cal L}_{2^n\times 2^n}$ such that in vector form, BN (\ref{2.1.1}) can be expressed by
\begin{align}\label{2.1.4}
x(t+1)=Mx(t),
\end{align}
where $M=M_1*M_2*\cdots*M_n$ is called the structure matrix of BN (\ref{2.1.1}).
\end{itemize}
Equation (\ref{2.1.4}) is called the algebraic state space representation (ASSR) of BN (\ref{2.1.1}).
\end{prp}

Recall equation (\ref{2.1.3}), $f$ is uniquely determined by the first (equivalently, second) row of $M_f$.
Hence, we may use one row (as a convention: we always choose the first row), denoted by $V_f$, to represent a Boolean function.

\begin{exa}\label{e2.1.2}
Consider $f(x_1,x_2,x_3)=(x_1\wedge x_3)\vee x_2$. It is easy to calculate its ASSR as
$$
\begin{array}{ccl}
f(x)&=&M_{\vee}M_{\wedge}x_1x_3x_2\\
~&=&M_{\vee}M_{\wedge}x_1W_{[2,2]}x_2x_3\\
~&=&M_{\vee}M_{\wedge}(I_2\otimes W_{[2,2]})x_1x_2x_3\\
~&=&M_fx,
\end{array}
$$
where $x=\ltimes_{i=1}^3x_i$,
$$
\begin{array}{ccl}
M_f&=&M_{\vee}M_{\wedge}(I_2\otimes W_{[2,2]})\\
~&=&\d_2[1,1,1,2,1,1,2,2].
\end{array}
$$
Hence,
\begin{align}\label{2.1.5}
V_f=[1,1,1,0,1,1,0,0].
\end{align}
\end{exa}

\begin{rem}\label{r2.1.3}
\begin{itemize}
\item[(i)] Using above notation (\ref{2.1.5}), for an $n$-node BN, there are $2^n$ states, hence the number of the Boolean functions is $2^{2^n}$, which is much larger than the number of states.
\item[(ii)] A Boolean function can be considered as an index function for a subset of states.
For instance, (\ref{2.1.5}) can be considered as a set of states $S=\d_8\{1,2,3,5,6\}$, which satisfy certain property characterized by $f$ (here the property is ``$f(x_1,x_2,x_3)$ is true", i.e. $f(x_1,x_2,x_3)=\delta_2^1$). Hence, $f$ can be considered as an index function of $S$.
\item[(iii)] Each Boolean vector $V\in {\cal B}^{2^n}$ represents a Boolean function $f(x_1,x_2,\cdots,x_n)$. In fact, they are one-to-one correspondence.
\end{itemize}
\end{rem}

\begin{dfn}\label{d2.1.301} Let $f(X)\in {\cal X}^*$ be a Boolean function. The support of $f$, denoted by $\supp(f)$, is defined as
$$
supp(f)=\{X\;|\; f(X)\neq 0\}=\{x\;|\;f(x)\neq \d_2^2\}.
$$
\end{dfn}

\begin{dfn}\label{d2.1.4} The state space of BN (\ref{2.1.1}), denoted by ${\cal X}$, is defined as
\begin{align}\label{2.1.6}
{\cal X}:=\{(X_1,X_2,\cdots,X_n)\;|\;X_i\in {\cal D},\;i=1,2,\cdots,n\}.
\end{align}
Equivalently, equation (\ref{2.1.6}) can be expressed by its vector form
\begin{align}\label{2.1.7}
{\cal X}:=\{x\;|\;x\in \D_{2^n}\}.
\end{align}
\end{dfn}

\begin{rem}\label{r2.1.5} Precisely speaking, (\ref{2.1.7}) should be expressed by
$$
\vec{\cal X}:=\{x\;|\;x\in \D_{2^n}\}.
$$
To avoid notational mess, ${\cal X}$ is used for both state space and the vector form of state space.
\end{rem}

\begin{dfn}\label{d2.1.6} The dual space of BN (\ref{2.1.1})'s state space, denoted by ${\cal X}^*$, is the set of Boolean functions. That is
\begin{align}\label{2.1.8}
{\cal X}^*:=\{Z\;|\; Z:{\cal X}\ra {\cal D}\}.
\end{align}
Equivalently, $Z$ can be expressed by its vector form $V_Z\in {\cal B}^{2^n}$, that is
\begin{align}\label{2.1.9}
{\cal X}^*:=\{V_Z\;|\;V_Z\in {\cal B}^{2^n}\}.
\end{align}
\end{dfn}

\begin{rem}\label{r2.1.7}
\begin{itemize}
\item[(i)] According to \cite{che10} and the following literature, all the logical functions of $\{x_1,x_2,\cdots,x_n\}$, denoted by
$$
{\cal F}_{\ell}\{x_1,x_2,\cdots,x_n\},
$$
is called the state space of BN (\ref{2.1.1}). This ``used" definition is, in certain sense, confusing. In this paper, the state space of BN (\ref{2.1.1}), denoted by ${\cal X}$, is defined as equation (\ref{2.1.6}) (or equivalently, equation (\ref{2.1.7})), and ${\cal F}_{\ell}\{x_1,x_2,\cdots,x_n\}={\cal X}^*$ is the dual (state) space. The new definitions clarify ``state space" and its ``dual space".
\item[(ii)] Since all the ``elements" in ${\cal X}^*$ are logical functions, meanwhile $x_i$, $i=1,2,\cdots,n$, are also logical functions. Hence
\begin{align}\label{2.1.10}
{\cal X}\subset {\cal X}^*.
\end{align}
\item[(iii)] There is no one-to-one correspondence between ${\cal X}$ and ${\cal X}^*$. In fact,
$|{\cal X}|=2^n$ and $|{\cal X}^*|=2^{2^n}$.
\item[(iv)] A set of Boolean functions ${\cal Z}=\{z_1,z_2,\cdots,z_s\}\subset {\cal X}^*$ can be expressed by
$$
z_i=G_ix,\quad i=1,2,\cdots,s,
$$
where $G_i\in {\cal L}_{2\times 2^n}$.\\
Setting $z=\ltimes_{i=1}^sz_i$, we have
\begin{align}\label{2.1.11}
z=Gx,
\end{align}
where $G=G_1*G_2*\cdots*G_s\in {\cal L}_{2^s\times 2^n}$.\\ ${\cal Z}^*:={\cal F}_{\ell}\{z_1,z_2,\cdots,z_s\}$ is called a subspace of state space (\cite{che10}). Precisely speaking, it is a subspace of the dual space.
\item[(v)] Assume $s=n$ and $G$ is non-singular, then
$$
{\cal Z}={\cal X},
$$
and equation (\ref{2.1.11}) is a coordinate change.
\item[(vi)]
Consider $\{z_1,z_2,\cdots,z_s\}\subset {\cal X}^*$. If there exists another set of logical functions
$\{z_{s+1},z_{s+2},\cdots,z_{n}\}\subset {\cal X}^*$, such that
$${\cal Z}^*={\cal F}_{\ell}\{z_1,z_2,\cdots,z_n\}={\cal X}^*,$$
then ${\cal Z}_0^*={\cal F}_{\ell}\{z_1,z_2,\cdots,z_s\}$ is a subspace of ${\cal X}^*$. That is,
$$
{\cal Z}_0^*\subset {\cal X}^*.
$$
It is called a regular subspace (\cite{che10}).
\end{itemize}
\end{rem}
The following result can be used to verify whether a given ${\cal Z}_0^*={\cal F}_{\ell}\{z_1,z_2,\cdots,z_s\}$ is a regular subspace.
\begin{prp}\label{p2.1.8} (\cite{che10}) Assume ${\cal Z}_0^*={\cal F}_{\ell}\{z_1,z_2,\cdots,z_s\}$ has its structure matrix $M_0$. That is, $z_0=M_0x$, where $M_0\in {\cal L}_{2^s\times 2^n}$. Then ${\cal Z}_0^*$ is a regular subspace, if and only if, $M_0$ has $2^{n-s}$ columns equal $\d_{2^s}^i$, $i=1,2,\cdots,2^s$.
That is,
\begin{align}\label{2.1.12}
\left|\left\{j\;|\;\Col_j(M_0)=\d_{2^s}^i\right\}\right|=2^{n-s},\quad i\in[1,2^s].
\end{align}
\end{prp}

\section{Exploring Hidden Structure of BN}

\subsection{Dual Network}

Consider BN (\ref{2.1.1}). Let ${\cal X}^*=\{z_1,z_2,\cdots,z_{2^{2^n}}\}$ be the dual space of state space ${\cal X}$. Since each $z_i\in {\cal X}^*$ is a Boolean function of ${\cal X}$,
$z_i$ can be expressed by
\begin{align}\label{3.1.0}
z_i=G_ix,\quad i\in [1,2^{2^n}],
\end{align}
where $G_i\in {\cal L}_{2\times 2^n}$ is the structure matrix of $z_i$.

\begin{dfn}\label{d3.1.0} Let $z\in {\cal X}^*$ be a Boolean function with its structure matrix $G_z$. Assume the first row of $G_z$ is
$$
\Row_1(G_z)=[\a_1,\a_2,\cdots,\a_{2^n}]\in {\cal B}^{2^n},
$$
which is called the structure vector of $z$.

Then the vector form of $z$, denoted by $\vec{z}$, is
\begin{align}\label{3.1.0.1}
\vec{z}=\d_{2^{2^n}}^i,
\end{align}
where
$$
i=\a_1(2^{2^n-1})+\a_2(2^{2^n-2})+\cdots+\a_{2^n-1}(2)+\a_{2^n}+1.
$$
\end{dfn}

Consequently, for $i\in[1,2^{2^n}]$, one has
\begin{align}\label{3.1.1}
\begin{array}{ccl}
z_i(t+1)=G_ix(t+1)=G_iMx(t)=z_{\phi(i)}(t)\in {\cal X}^*,
\end{array}
\end{align}
where the structure matrix of $z_{\phi(i)}$ is $G_iM$.

In vector form, BN (\ref{3.1.1}) is expressed by
\begin{align}\label{3.1.2}
z(t+1)=\d_{2^{2^n}}\left[\phi(1),\phi(2),\cdots,\phi(2^{2^n})\right]z(t),
\end{align}
which is the dynamic equation of logical functions.

\begin{dfn}\label{d3.1.1} Consider BN (\ref{2.1.1}) with its ASSR (\ref{2.1.4}).\\
The dynamic equation (\ref{3.1.2}) is called the dual BN of BN (\ref{2.1.1}).
\end{dfn}

Here we provide an example to demonstrate this.

\begin{exa}\label{e3.1.2}
Consider the following $2$-node BN
\begin{align}\label{3.1.8}
\begin{cases}
x_1(t+1)=x_2(t),\\
x_2(t+1)=\neg(x_1(t)\vee x_2(t)).
\end{cases}
\end{align}
The ASSR is
\begin{align}\label{3.1.9}
x(t+1)=Mx(t),
\end{align}
where $M=\d_4[2,3,2,4]$.

The state transition graph of BN (\ref{3.1.8}) is shown in Fig. \ref{Fig.3.1}.

\begin{figure}
\centering
\setlength{\unitlength}{0.7 cm}
\begin{picture}(12,6)\thicklines
\put(6,1.5){\oval(2,1)}
\put(2,4.5){\oval(2,1)}
\put(6,4.5){\oval(2,1)}
\put(10,4.5){\oval(2,1)}
\put(3,4.5){\vector(1,0){2}}
\put(7,4.75){\vector(1,0){2}}
\put(9,4.25){\vector(-1,0){2}}
\put(7.5,1.5){\oval(2,0.5)[r]}
\put(7,1.75){\line(1,0){0.5}}
\put(7.5,1.25){\vector(-1,0){0.5}}
\put(5.7,1.3){$\d_4^4$}
\put(1.7,4.3){$\d_4^1$}
\put(5.7,4.3){$\d_4^2$}
\put(9.7,4.3){$\d_4^3$}
\end{picture}
\caption{State Transition Graph of BN (\ref{3.1.8})\label{Fig.3.1}}
\end{figure}
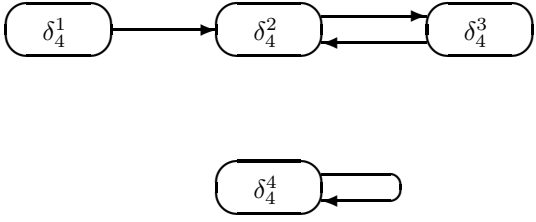

In this example, there are totally $2^{2^2}=16$ logical functions for state space ${\cal X}=\Delta_{2^2}$. Using their structure vectors, we can arrange them in alphabetic form as
$$
\begin{array}{cc}
V_{z_1}:=[0,0,0,0],&z_1:=\d_{16}^1;\\
V_{z_2}:=[0,0,0,1],&z_2:=\d_{16}^2;\\
\quad\quad\quad\quad\quad\quad\vdots\\
V_{z_{16}}:=[1,1,1,1],&z_{16}:=\d_{16}^{16}.\\
\end{array}
$$
From equation (\ref{3.1.2}), one has
\begin{align}
V_{z(t+1)}=V_{z(t)}M.
\end{align}
A straightforward computation shows that
\begin{align}\label{3.1.10}
\begin{bmatrix}
0&0&0&0\\
0&0&0&1\\
0&0&1&0\\
0&0&1&1\\
0&1&0&0\\
0&1&0&1\\
0&1&1&0\\
0&1&1&1\\
1&0&0&0\\
1&0&0&1\\
1&0&1&0\\
1&0&1&1\\
1&1&0&0\\
1&1&0&1\\
1&1&1&0\\
1&1&1&1\\
\end{bmatrix}M=
\begin{bmatrix}
0&0&0&0\\
0&0&0&1\\
0&1&0&0\\
0&1&0&1\\
1&0&1&0\\
1&0&1&1\\
1&1&1&0\\
1&1&1&1\\
0&0&0&0\\
0&0&0&1\\
0&1&0&0\\
0&1&0&1\\
1&0&1&0\\
1&0&1&1\\
1&1&1&0\\
1&1&1&1\\
\end{bmatrix}
=\begin{bmatrix}
V_{z_1}\\
V_{z_2}\\
V_{z_5}\\
V_{z_6}\\
V_{z_{11}}\\
V_{z_{12}}\\
V_{z_{15}}\\
V_{z_{16}}\\
V_{z_{1}}\\
V_{z_{2}}\\
V_{z_{5}}\\
V_{z_{6}}\\
V_{z_{11}}\\
V_{z_{12}}\\
V_{z_{15}}\\
V_{z_{16}}\\
\end{bmatrix}.
\end{align}
Setting $z_i:=\d_{16}^i,~i=1,2,\cdots,16$, one can obtain the ASSR of the dual BN as follows:
\begin{align}\label{3.1.11}
z(t+1)=M^*z(t),
\end{align}
where
$$
M^*=\d_{16}[1,2,5,6,11,12,15,16,1,2,5,6,11,12,15,16].
$$
Fig. \ref{Fig.3.2} Shows the state transition graph of dual BN (\ref{3.1.11}).
Inside each oval is the structure vector of the corresponding logical function.

\vskip 5mm

\begin{figure}
\centering
\setlength{\unitlength}{0.8 cm}
\begin{picture}(8,19)\thicklines
\put(2,1.5){\oval(2,1)}
\put(6,1.5){\oval(2,1)}
\put(2,3.5){\oval(2,1)}
\put(6,3.5){\oval(2,1)}
\put(2,5.5){\oval(2,1)}
\put(6,5.5){\oval(2,1)}
\put(2,7.5){\oval(2,1)}
\put(6,7.5){\oval(2,1)}
\put(2,11.5){\oval(2,1)}
\put(6,11.5){\oval(2,1)}
\put(2,13.5){\oval(2,1)}
\put(6,13.5){\oval(2,1)}
\put(2,15.5){\oval(2,1)}
\put(6,15.5){\oval(2,1)}
\put(2,17.5){\oval(2,1)}
\put(6,17.5){\oval(2,1)}
\put(3,1.75){\vector(1,0){2}}
\put(5,1.25){\vector(-1,0){2}}
\put(2,3){\vector(0,-1){1}}
\put(6,3){\vector(0,-1){1}}
\put(3,5.75){\vector(1,0){2}}
\put(5,5.25){\vector(-1,0){2}}
\put(2,7){\vector(0,-1){1}}
\put(6,7){\vector(0,-1){1}}
\put(3,11.5){\vector(1,0){2}}
\put(3,13.5){\vector(1,0){2}}
\put(3,15.5){\vector(1,0){2}}
\put(3,17.5){\vector(1,0){2}}
\put(7.5,11.5){\oval(2,0.5)[r]}
\put(7,11.75){\line(1,0){0.5}}
\put(7.5,11.25){\vector(-1,0){0.5}}
\put(7.5,13.5){\oval(2,0.5)[r]}
\put(7,13.75){\line(1,0){0.5}}
\put(7.5,13.25){\vector(-1,0){0.5}}
\put(7.5,15.5){\oval(2,0.5)[r]}
\put(7,15.75){\line(1,0){0.5}}
\put(7.5,15.25){\vector(-1,0){0.5}}
\put(7.5,17.5){\oval(2,0.5)[r]}
\put(7,17.75){\line(1,0){0.5}}
\put(7.5,17.25){\vector(-1,0){0.5}}
\put(1.1,1.3){$(1,0,1,1)$}
\put(5.1,1.3){$(0,1,0,1)$}
\put(1.1,3.3){$(1,1,0,1)$}
\put(5.1,3.3){$(0,0,1,1)$}
\put(1.1,5.3){$(1,0,1,0)$}
\put(5.1,5.3){$(0,1,0,0)$}
\put(1.1,7.3){$(1,1,0,0)$}
\put(5.1,7.3){$(0,0,1,0)$}
\put(1.1,11.3){$(1,0,0,0)$}
\put(5.1,11.3){$(0,0,0,0)$}
\put(1.1,13.3){$(1,0,0,1)$}
\put(5.1,13.3){$(0,0,0,1)$}
\put(1.1,15.3){$(0,1,1,0)$}
\put(5.1,15.3){$(1,1,1,0)$}
\put(1.1,17.3){$(0,1,1,1)$}
\put(5.1,17.3){$(1,1,1,1)$}
\end{picture}
\caption{Transition Graph of Dual BN (\ref{3.1.11})\label{Fig.3.2}}
\end{figure}
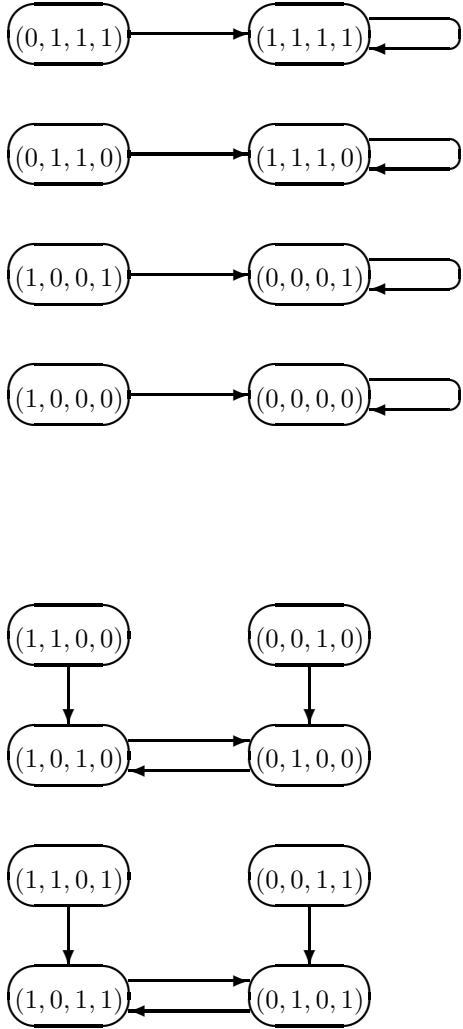

\end{exa}


\subsection{Dual Attractor}

\begin{dfn}\label{d3.2.3}
Let $f(x)\in {\cal X}^*$ be a Boolean function with its structure matrix
$G_f$, that is,
\begin{align}\label{3.2.12}
f(x)=G_fx,
\end{align}
\begin{itemize}
\item[(i)] $f(x)$ is called a dual fixed point, if
\begin{align}\label{3.2.13}
G_fM=G_f.
\end{align}
\item[(ii)] $(f_1(x),f_2(x),\cdots,f_{\ell+1}(x)=f_1(x))$ is called a dual cycle with length $\ell$, if their structure matrices $G_i=G_{f_i}$, $i=1,2,\cdots,\ell+1$ satisfy:
\begin{enumerate}
    \item[(a)]
$$
G_i\neq G_j,\quad i\neq j,~1\leq i,j\leq \ell;
$$
\item[(b)]
$$
G_1=G_{\ell+1};
$$
\item[(c)]
\begin{align}\label{3.2.14}
G_{j+1}=G_jM,\quad j=1,2,\cdots,\ell.
\end{align}
\end{enumerate}
\end{itemize}
\end{dfn}

\begin{rem}\label{r3.2.4}
\begin{itemize}
\item[(i)] The dual fixed points and dual cycles are together called dual attractors, because they are the attractors of the dual BN.  Since a BN has finite (precisely, $2^{2^n}$) logical functions, each logical function will converge to an attractor.
\item[(ii)] In  equation (\ref{3.2.14}), the structure matrix of $f$, i.e., $G_f$ or $G_j$, can be replaced by their structure vectors $V_f$ or $V_{f_j}$ as shown in equation (\ref{3.1.10}).

\item[(iii)] Similarly to attractors in BNs, the basin of a set of dual attractors $A^*\subset {\cal X}^*$, is defined as
\begin{align}\label{3.2.15}
B_{A^*}:=\{z\in {\cal X}^*\;|\; \exists N\in{\mathbb N}, s.t.~zM^s\in A^*,\;s\geq N\}.
\end{align}
\end{itemize}
\end{rem}

\begin{exa}\label{e3.2.5} Recall Example \ref{e3.1.2}.
From Fig. \ref{Fig.3.2}, one sees easily that there are $4$ dual fixed points
$$
\begin{array}{ll}
V_{z_1}=(0,0,0,0);&V_{z_2}=(0,0,0,1);\\
V_{z_{15}}=(1,1,1,0);&V_{z_{16}}=(1,1,1,1);
\end{array}
$$
and $2$ dual cycles
$$
\begin{array}{l}
V_{z_{11}}=(1,0,1,0) \leftrightharpoons V_{z_{5}}=(0,1,0,0);\\
V_{z_{12}}=(1,0,1,1) \leftrightharpoons V_{z_{6}}=(0,1,0,1).\\
\end{array}
$$
The basin of each attractor is also obvious in Fig. \ref{Fig.3.2}.
\end{exa}

\subsection{Invariant Subspace}

Consider BN (\ref{2.1.1}) with its ASSR (\ref{2.1.4}).  Starting from any $z_0\in {\cal X}^*$, we can construct a subset of ${\cal X}^*$ as
\begin{align}\label{3.3.1}
\{z(0)=z_0,z(1)=M^*z(0),\cdots,z(N)=M^*z(N-1)\},
\end{align}
where $M^*$ is the structure matrix of the dual BN as shown in equation (\ref{3.1.11}).

Then \cite{chepr} concludes that
\begin{align}\label{3.3.2}
{\cal Z}_0:=\{z(0),z(1),\cdots,z(N)\}
\end{align}
is $M$-invariant, if
\begin{align}\label{3.3.3}
\{z\;|\;z=M^*z(i),i=0,1,
\cdots, N\}\subset {\cal Z}_0.
\end{align}
If $N$ is the smallest natural number that satisfies equation (\ref{3.3.3}), then the corresponding ${\cal Z}_0$ is the
smallest $M$-invariant subspace containing $z_0$.

In a similar way, we can also construct a (smallest) $M$-invariant subspace ${\cal V}$ containing a subset $
{\cal Z}\subset {\cal X}^*
$. It is obvious that ${\cal V}$ is a subspace of ${\cal X}^*$. Furthermore, ${\cal V}$ can be considered as a subspace of ${\cal X}$, only when ${\cal V}$ is a regular subspace.

Since ${\cal Z}^*\subset {\cal X}^*$, there exists its structure matrix $M_z$, such that
$$
z=\vec{{\cal Z}^*}=M_zx,
$$
where $M_z=G_{j_1}*G_{j_2}*\cdots*G_{j_{|{\cal Z}^*|}}$, $G_{j_i}$ is the structure matrix of $z_{j_i}\in {\cal Z}^*$, $z=\ltimes_{i=1}^{|{\cal Z}^*|}z_{j_i}$, $x=\ltimes_{j=1}^nx_j$.
Then (\cite{chepr}) there exists a logical matrix $H\in{\cal L}_{2^{|{\cal Z}^*|}\times 2^{|{\cal Z}^*|}}$ , such that
\begin{align}\label{3.3.4}
M_zM=HM_z.
\end{align}
It follows that
\begin{align}\label{3.3.5}
\begin{array}{ccl}
z(t+1)&=&M_zx(t+1)=M_zMx(t)\\
~&=&HM_zx(t)=Hz(t).
\end{array}
\end{align}

According to \cite{chepr}, BN (\ref{3.3.5}) is a minimum realization of BN (\ref{2.1.1}) involving $z_0$.

Consider each dual attractor set $A^*_i$ of BN  (\ref{2.1.1}), the following result is obvious:

\begin{prp}\label{p3.3.1} Let $A^*_i,~i=1,2,\cdots,s$ be the sets of dual attractions of BN (\ref{2.1.1}), and the basin of attraction for $A^*_i$ are $B^*_i$ $i=1,2,\cdots,s$. Let $C^*_i:=A^*_i\cup B^*_i$, then $C^*_i$, $i=1,2,\cdots,s$ are $M$-invariant subspaces. Moreover, they form a partition of the dual space ${\cal X}^*$. That is,
\begin{align}\label{3.3.6}
{\cal X}^*=\bigcup_{i=1}^sC^*_i,
\end{align}
and $C^*_i\cap C^*_j=\emptyset$ for $i\neq j$.
\end{prp}

Observing Example \ref{e3.1.2}, one can easily obtain the following result.

\begin{prp}\label{p3.5} Consider BN (\ref{2.1.1}).
\begin{itemize}
\item[(i)] Assume $z\in {\cal F}_{\ell}\{x_1,x_2,\cdots,x_n\}$ is a dual fixed point, then
$\neg z$ is also a dual fixed point.
\item[(ii)] Assume $(z_0,z_1,\cdots,z_{\ell}=z_0)\subset {\cal F}_{\ell}\{x_1,x_2,\cdots,x_n\}$ is a dual cycle, then $(\neg z_0,\neg z_1,\cdots, \neg z_{\ell})$ is also a dual cycle.
\item[(iii)] If $B^*_i$ is the basin of attraction for attractor set $A^*_i$, then
$\neg B^*_i$ is the basin of attraction for attractor set $\neg A^*_i$.
\end{itemize}
\end{prp}

{\bf Proof:}
Assume $z$ is a dual fixed point whose structure matrix is $G_z$. Then it is straightforward that $$G_{\neg z}=G_{\neg}G_{z},$$ where $G_{\neg}=\delta_2[2,1]$ is the structure matrix of negation. \\Since $z$ is a dual fixed point, we have $G_zM=G_z$, then $$G_{\neg z}M=(G_{\neg}G_z)M=G_{\neg}(G_zM)=G_{\neg}G_z=G_{\neg z},$$
which means $\neg z$ is a dual fixed point. The proofs of other results are similar.
\hfill $\Box$

\section{Attractor vs Dual Attractor}

The canonical form of a BN has been discussed in \cite{for13,liu19}. In the following, we use the framework provided by \cite{liu19}.

A matrix is called a cyclic matrix, if it can be expressed by
$$
A=\d_k[2,3,\cdots,k,1].
$$
A matrix $B$ is called a Nilpotent matrix, if there is an $s\in {\mathbb N}$ such that $B^s= {\bf 0}$ (\cite{hor85}).

\begin{prp}\label{p4.1} Consider BN (\ref{2.1.1}) with its ASSR (\ref{2.1.4}).
There exists a coordinate change $\tilde{x}=Tx$, such that under $\tilde{x}$, equation (\ref{2.1.4}) becomes
\begin{align}\label{4.1}
\tilde{x}(t+1)=\tilde{M}\tilde{x}(t)
 :=\begin{bmatrix}
C_1&{\bf 0}&\cdots&{\bf 0}\\
{\bf 0}&C_2&\cdots&{\bf 0}\\
\vdots&\vdots&\ddots&\vdots\\
{\bf 0}&{\bf 0}&\cdots&C_s\\
\end{bmatrix}\tilde{x}(t)
,
\end{align}
where
$
C_i=\begin{bmatrix}
A_i&E_i\\
{\bf 0}&B_i
\end{bmatrix},~i=1,2,\cdots,s
$, $A_i$ is a cyclic matrix and $B_i$ is a Nilpotent matrix.
\end{prp}

In fact, $C_i$ corresponds to cycle $A_i$ and its basin of attraction $B_i$.
Precisely speaking, if the $i$ th subnetwork of (\ref{4.1}) corresponding to $C_i$ is expressed into block-wise form as
$$
\begin{bmatrix}
x^i_1(t+1)\\
x^i_2(t+1)
\end{bmatrix}=
\begin{bmatrix}
A_i&E_i\\
{\bf 0}&B_i
\end{bmatrix}
\begin{bmatrix}
x^i_1(t)\\
x^i_2(t)
\end{bmatrix},
$$
then the set $\{x^i_1\}$ forms the $i$ th cycle and  the set $\{x^i_2\}$ forms its basin of attraction.

Assume $C_i\in {\cal M}_{r_i\times r_i}$, $i=1,2,\cdots,s$, then
$$
\dsum_{i=1}^s r_i=2^n.
$$

Since each BN has its canonical form, in the following we assume $x$ itself is a ``canonical" coordinate frame such that the BN under $x$ is in the canonical form.


Recall canonical form (\ref{4.1}). Assume the states are partitioned into
$$
{\cal X}=\bigcup_{i=1}^s{\cal X}_i,
$$
where ${\cal X}_i:=\{\delta_{2^n}^j\;|\;\tilde{M}\delta_{2^n}^j~\mbox{includes a column of}~C_i\}$ corresponds to the states in the $i$-th block in equation (\ref{4.1}).

Then we have the following result.
\begin{prp}\label{p4.3}
Let ${\cal X}_i^*=\{f\;|\; f\in {\cal X}^*,\supp(f)\subset {\cal X}_i\}$, $i\in [1,s]$. Then  ${\cal X}_i^*$ are $\tilde{M}$-invariant subspaces.
\end{prp}
{\bf Proof:}
We can rewrite ${\cal X}^*_i$ as
\begin{align*}
    {\cal X}^*_i=\{f\;|\;f\in {\cal X}^*,~f(x)=\delta_2^1 \Rightarrow x\in {\cal X}_i\},
\end{align*}
that is, if $V_f(j)=1~\mbox{then}~\delta_{2^n}^j\in {\cal X}_i,$ where $V_f$ is the structure vector of $f$.

It is obvious that for $\forall f\in{\cal X}_i^*$, if $[V_f\tilde{M}](j)=1~\mbox{then}~\delta_{2^n}^j\in {\cal X}_i$, which means that $f'\in{\cal X}_i^*$, where $V_{f'}=V_f\tilde{M}$. Thus, ${\cal X}_i^*$ is an $\tilde{M}$-invariant subspace.
\hfill $\Box$

\begin{exa} \label{e4.4} Consider BN
\begin{align}\label{4.2}
\left\{
\begin{array}{lll}
Z_1(t+1)&=&Z_1(t)\vee \{\neg Z_1(t) \wedge [(\neg Z_2(t))\wedge Z_3(t)]\},\\
Z_2(t+1)&=&[\neg Z_1(t)\wedge(Z_2(t)\lra Z_3(t))]\\
~&~&\vee[\neg(Z_2(t)\wedge Z_3(t))],\\
Z_3(t+1)&=&(Z_1(t)\wedge Z_3(t)\vee (\neg Z_1(t)\wedge \neg Z_2(t)).
\end{array}
\right.
\end{align}

Its ASSR is calculated as
\begin{align}\label{4.3}
z(t+1)=Mz(t),
\end{align}
where $z(t)=\prod_{i=1}^3z_i(t)$, $
M=\d_8[3,2,1,2,6,8,3,5]$.\\
Its state transition graph is depicted in Fig. \ref{Fig.4.1}.

\begin{figure}
\centering
\setlength{\unitlength}{0.7 cm}
\begin{picture}(12,11)\thicklines
\put(2,1.5){\oval(2,1)}
\put(6,1.5){\oval(2,1)}
\put(10,1.5){\oval(2,1)}
\put(2,4.5){\oval(2,1)}
\put(6,4.5){\oval(2,1)}
\put(2,7.5){\oval(2,1)}
\put(6,7.5){\oval(2,1)}
\put(4,9.5){\oval(2,1)}
\put(3,1.5){\vector(1,0){2}}
\put(7,1.75){\vector(1,0){2}}
\put(9,1.25){\vector(-1,0){2}}
\put(3,4.5){\vector(1,0){2}}
\put(3,7.5){\vector(1,0){2}}
\put(6,8){\vector(-1,1){1}}
\put(3,9){\vector(-1,-1){1}}
\put(7.5,4.5){\oval(2,0.5)[r]}
\put(7,4.75){\line(1,0){0.5}}
\put(7.5,4.25){\vector(-1,0){0.5}}
\put(1.7,1.3){$\d_8^7$}
\put(1.7,4.3){$\d_8^4$}
\put(1.7,7.3){$\d_8^5$}
\put(3.7,9.3){$\d_8^4$}
\put(5.7,1.3){$\d_8^3$}
\put(5.7,4.3){$\d_8^2$}
\put(5.7,7.3){$\d_8^6$}
\put(9.7,1.3){$\d_8^1$}
\end{picture}
\caption{Transition Graph of BN (\ref{4.2})\label{Fig.4.1}}
\end{figure}
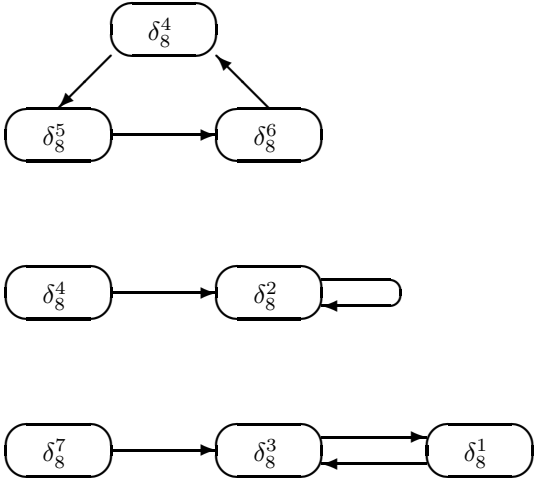

According to the state transition graph Fig. \ref{Fig.4.1}, a coordinate transformation can be obtained as in \cite{liu19}:
\begin{align}\label{4.4}
x=Tz,
\end{align}
where $T=\d_8[1,4,2,5,6,7,3,8]$.\\
Under coordinates $x$, system (\ref{4.3}) becomes
\begin{align}\label{4.5}
x(t+1)=\tilde{M}x(t),
\end{align}
where
$$
\begin{array}{ccl}
\tilde{M}&=&T*M*T^\mathrm{T}\\
~&=&\d_8[2,1,2,4,4,7,8,6].
\end{array}
$$
Now BN (\ref{4.5}) is in canonical form and the structure matrix is
$$
\tilde{M}=\diag(C_1,C_2,C_3),
$$
where
$
C_i=\begin{bmatrix}
A_i&E_i\\
{\bf 0}&B_i
\end{bmatrix},\quad i=1,2,3
$
with
$$
\begin{array}{lll}
A_1=\d_2[2,1],&E_1=\d_2^2,&B_1=0;\\
A_2=1,&E_2=1,&B_2=0;\\
A_3=\d_3[2,3,1].&~&~
\end{array}
$$
The dual subspaces are
$$
\begin{array}{lll}
{\cal X}^*_1=\{f(x)\in {\cal X}^*\;|\;\supp(f)\subset {\cal X}_1\},\\
{\cal X}^*_2=\{f(x)\in {\cal X}^*\;|\;\supp(f)\subset {\cal X}_2\},\\
{\cal X}^*_3=\{f(x)\in {\cal X}^*\;|\;\supp(f)\subset {\cal X}_3\},\\
\end{array}
$$
where $c_i\in {\cal D}$, $i=1,2,\cdots,8$.

Finally, we consider topology of the dual space. Since ${\cal X}^*_i$ are invariant, attractors within ${\cal X}^*_i$ with their basins of attraction form a partition of ${\cal X}^*_i$. The topological structures of ${\cal X}^*_i$ are discussed one by one as follows:

\begin{itemize}
\item[(i)] Consider ${\cal X}^*_1$: Let $f\in {\cal X}^*_1$. Then $V_f=(c_1,c_2,c_3,0,0,0,0,0)$. Using $(c_1,c_2,c_3)$ to represent $f$,  we have dual structure on ${\cal X}^*_1$ as given in Fig. \ref{Fig.4.2}.
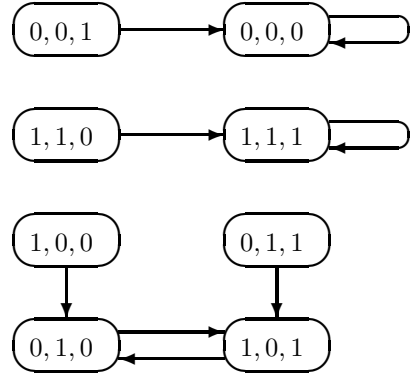
\begin{figure}
\centering
\setlength{\unitlength}{0.7 cm}
\begin{picture}(9,9)\thicklines
\put(2,1.5){\oval(2,1)}
\put(6,1.5){\oval(2,1)}
\put(2,3.5){\oval(2,1)}
\put(6,3.5){\oval(2,1)}
\put(2,5.5){\oval(2,1)}
\put(6,5.5){\oval(2,1)}
\put(2,7.5){\oval(2,1)}
\put(6,7.5){\oval(2,1)}
\put(3,1.75){\vector(1,0){2}}
\put(5,1.25){\vector(-1,0){2}}
\put(2,3){\vector(0,-1){1}}
\put(6,3){\vector(0,-1){1}}
\put(3,5.5){\vector(1,0){2}}
\put(3,7.5){\vector(1,0){2}}
\put(7.5,5.5){\oval(2,0.5)[r]}
\put(7,5.75){\line(1,0){0.5}}
\put(7.5,5.25){\vector(-1,0){0.5}}
\put(7.5,7.5){\oval(2,0.5)[r]}
\put(7,7.75){\line(1,0){0.5}}
\put(7.5,7.25){\vector(-1,0){0.5}}
\put(1.3,1.3){$0,1,0$}
\put(1.3,3.3){$1,0,0$}
\put(1.3,5.3){$1,1,0$}
\put(1.3,7.3){$0,0,1$}
\put(5.3,1.3){$1,0,1$}
\put(5.3,3.3){$0,1,1$}
\put(5.3,5.3){$1,1,1$}
\put(5.3,7.3){$0,0,0$}
\end{picture}
\caption{Dual Structure on ${\cal X}^*_1$ \label{Fig.4.2}}
\end{figure}

\item[(ii)]
Consider ${\cal X}^*_2$: Let $f\in {\cal X}^*_2$. Then $V_f=(0,0,0,c_4,c_5,0,0,0)$. Using $(c_4,c_5)$ to represent $f$,  we have
dual structure on ${\cal X}^*_2$ as given in Fig. \ref{Fig.4.3}.

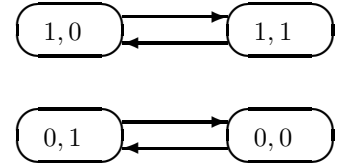
\begin{figure}
\centering
\setlength{\unitlength}{0.7 cm}
\begin{picture}(5,8)\thicklines
\put(2,1.5){\oval(2,1)}
\put(6,1.5){\oval(2,1)}
\put(2,3.5){\oval(2,1)}
\put(6,3.5){\oval(2,1)}
\put(3,1.75){\vector(1,0){2}}
\put(5,1.25){\vector(-1,0){2}}
\put(3,3.75){\vector(1,0){2}}
\put(5,3.25){\vector(-1,0){2}}
\put(1.5,1.3){$0,1$}
\put(1.5,3.3){$1,0$}
\put(5.5,1.3){$0,0$}
\put(5.5,3.3){$1,1$}
\end{picture}
\caption{Dual Structure on ${\cal X}^*_2$ \label{Fig.4.3}}
\end{figure}

\item[(iii)]
Finally, consider ${\cal X}^*_3$: Let $f\in {\cal X}^*_3$. Then $V_f=(0,0,0,0,0,c_6,c_7,c_8)$. $(c_6,c_7,c_8)$ to represent $f$, we have dual structure on ${\cal X}^*_3$ as given in Fig. \ref{Fig.4.4}.

\begin{figure}
\centering
\setlength{\unitlength}{0.7 cm}
\begin{picture}(7,9)\thicklines
\put(4,1.5){\oval(2,1)}
\put(2,3.5){\oval(2,1)}
\put(6,3.5){\oval(2,1)}
\put(2,5.5){\oval(2,1)}
\put(6,5.5){\oval(2,1)}
\put(3,3.5){\vector(1,0){2}}
\put(3.5,2){\vector(-1,1){1}}
\put(4.5,2){\vector(1,1){1}}
\put(3.5,5.5){\oval(2,0.5)[r]}
\put(3,5.75){\line(1,0){0.5}}
\put(3.5,5.25){\vector(-1,0){0.5}}
\put(7.5,5.5){\oval(2,0.5)[r]}
\put(7,5.75){\line(1,0){0.5}}
\put(7.5,5.25){\vector(-1,0){0.5}}
\put(3.3,1.3){$0,0,1$}
\put(1.3,3.3){$1,0,0$}
\put(5.3,3.3){$0,1,0$}
\put(1.3,5.3){$1,1,1$}
\put(5.3,5.3){$0,0,0$}
\end{picture}
\caption{Dual Structure on ${\cal X}^*_3$ \label{Fig.4.4}}
\end{figure}
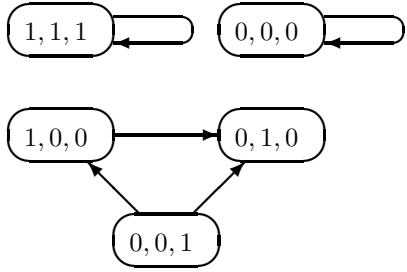

\end{itemize}

\end{exa}

\section{Hidden Order}

The topological structure on ${\cal X}^*={\cal F}_{\ell}$ is not obvious. Particularly, under the original coordinate frame, which is the natural one, it is even murky.

Observe the following example.

\begin{exa}\label{e5.1} Recall Example \ref{e4.4}. Consider the functions in ${\cal X}^*_1$. Back to the original coordinate frame, it is easy to calculate their structure vectors, see TABLE \ref{Tab.5.1}.

\vskip 2mm

\begin{table}
\centering \caption{Structure Vector of Functions in $D_1$}\label{Tab.5.1} 
\begin{tabular}{|c||c|}
\hline $(c_1,c_2,c_3)$& $V_z$\\
\hline
(0,0,0)&(0,0,0,0,0,0,0,0)\\
\hline
(0,0,1)&(0,1,0,0,0,0,0,0)\\
\hline
(0,1,0)&(0,0,0,1,0,0,0,0)\\
\hline
(0,1,1)&(0,1,0,1,0,0,0,0)\\
\hline
(1,0,0)&(1,0,0,0,0,0,0,0)\\
\hline
(1,0,1)&(1,1,0,0,0,0,0,0)\\
\hline
(1,1,0)&(1,0,0,1,0,0,0,0)\\
\hline
(1,1,1)&(1,1,0,1,0,0,0,0).\\
\hline
\end{tabular}
\end{table}

Similarly, for functions in ${\cal X}^*_2$, back to the original coordinate frame, their structure vectors are shown in TABLE \ref{Tab.5.2}.

\vskip 2mm

\begin{table}
\centering \caption{Structure Vector of Functions in $D_2$}\label{Tab.5.2} 
\begin{tabular}{|c||c|}
\hline $(c_4,c_5)$& $V_z$\\
\hline
(0,0)&(0,0,0,0,0,0,0,0)\\
\hline
(0,1)&(0,0,0,0,0,1,0,0)\\
\hline
(1,0)&(0,0,0,0,1,0,0,0)\\
\hline
(1,1)&(0,0,0,0,1,1,0,0)\\
\hline
\end{tabular}
\end{table}

For functions in ${\cal X}^*_3$, back to the original coordinate frame, their structure vectors are shown in TABLE \ref{Tab.5.3}.

\vskip 2mm

\begin{table}
\centering \caption{Structure Vector of Functions in $D_3$}\label{Tab.5.3} 
\begin{tabular}{|c||c|}
\hline $(c_1,c_2,c_3)$& $V_z$\\
\hline
(0,0,0)&(0,0,0,0,0,0,0,0)\\
\hline
(0,0,1)&(0,0,0,0,0,0,0,1)\\
\hline
(0,1,0)&(0,0,1,0,0,0,0,0)\\
\hline
(0,1,1)&(0,0,1,0,0,0,0,1)\\
\hline
(1,0,0)&(0,0,0,0,0,0,1,0)\\
\hline
(1,0,1)&(0,0,0,0,0,0,1,1)\\
\hline
(1,1,0)&(0,0,1,0,0,0,1,0)\\
\hline
(1,1,1)&(0,0,1,0,0,0,1,1).\\
\hline
\end{tabular}
\end{table}

\end{exa}

\begin{figure}
\centering
\setlength{\unitlength}{5 mm}
\begin{picture}(15,15)
\put(1,1){\framebox(4,13){}}
\put(6,1){\framebox(8,13){}}
\thicklines
\put(2,2){\framebox(2,2.5){$C_s$}}
\put(2,7){\framebox(2,2.5){$C_2$}}
\put(2,11){\framebox(2,2.5){$C_1$}}
\put(3,5.5){$\vdots$}
\put(7,2){\framebox(6,2.5){}}
\put(7,7){\framebox(6,2.5){}}
\put(7,11){\framebox(6,2.5){}}
\put(10,5.5){$\vdots$}
\put(7.2,3.8){$D_s$}
\put(8,2.5){\framebox(1.5,1){$D^1_s$}}
\put(11,2.5){\framebox(1.5,1){$D^{\xi_s}_s$}}
\put(9.8,3){$\cdots$}
\put(7.2,8.8){$D_2$}
\put(8,7.5){\framebox(1.5,1){$D^1_2$}}
\put(11,7.5){\framebox(1.5,1){$D^{\xi_2}_2$}}
\put(9.8,8){$\cdots$}
\put(7.2,12.8){$D_1$}
\put(8,11.5){\framebox(1.5,1){$D^1_1$}}
\put(11,11.5){\framebox(1.5,1){$D^{\xi_1}_1$}}
\put(9.8,12){$\cdots$}
\put(0.7,14.4){Attractors on ${\cal X}$}
\put(7.5,14.4){Attractors on ${\cal X}^*$}
\put(4,3.25){\vector(1,0){3}}
\put(4,8.25){\vector(1,0){3}}
\put(4,12.25){\vector(1,0){3}}
\end{picture}
\caption{Order vs Hidden Order of BN \label{Fig.5.1}}
\end{figure}
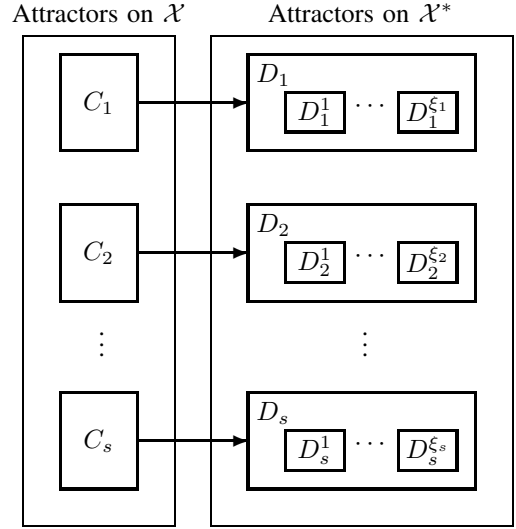

Combining Example \ref{e4.4} with Example \ref{e5.1}, one sees easily that finding the order determined by dual logical functions is not easy. Hence the order suggested by the topological structure of a dual BN (i.e., dynamics of logical functions) is called the hidden order. From Examples \ref{e4.4} and \ref{e5.1} one sees that the hidden order determines the behaviors of the BN. The relationship between attractors and dual attractors is depicted by Fig. \ref{Fig.5.1}.


\section{Boolean Algebra on ${\cal X}^*$}

Recall network (\ref{4.2}) in Example \ref{e4.4}, which has nodes $n=3$.
Hence, $|{\cal X}^*|=2^{2^3}=256$. But in Example \ref{e5.1} only ${\cal X}^*_i$, $i=1,2,3$, which involve only $20$ elements of ${\cal X}^*$,  have been investigated.
Are ${\cal X}^*_i$, $i=1,2,3$ enough to determine the topological structure of ${\cal X}^*$? The answer is ``yes". The reason is that the elements in  ${\cal X}^*$ are not independent. There is a Boolean algebra structure over ${\cal X}^*$. This section is devoted to the algebraic structure over ${\cal X}^*$.

\begin{dfn}\label{d.15.1} \cite{ros03} A Boolean algebra is ${\cal B}_A=\{\wedge, \vee, \neg, {\bf 1}, {\bf 0}\}$,  satisfying
\begin{itemize}
\item[(i)] commutative laws:
$$
\begin{cases}
x\vee y=y\vee x,\\
x\wedge y=y\wedge x.
\end{cases}
$$

\item[(ii)] Associative Laws:
$$
\begin{cases}
(x\vee y)\vee z=x\vee(y\vee z),\\
(x\wedge y)\wedge z=x\wedge(y\wedge z).\\
\end{cases}
$$

\item[(iii)] Distributive Laws:
$$
\begin{cases}
x\vee (y \wedge z)=(x\vee y) \wedge (x\vee z),\\
x\wedge (y\vee z)=(x\wedge y) \vee  (x\wedge z).\\
\end{cases}
$$
\item[(iv)] Identity Laws:
$$\begin{cases}
x\vee {\bf 0}=x,\\
x\wedge {\bf 1}=x.\end{cases}
$$

\item[(v)] Complement Laws:
$$\begin{cases}
x\vee \neg x={\bf 1},\\
x\wedge \neg x={\bf 0}.\end{cases}
$$
\end{itemize}
\end{dfn}

\begin{dfn}\label{d.15.2}
Let $f,g\in {\cal X}^*$ with their structure vectors $V_f$ and $V_g$, respectively. Then
\begin{itemize}
\item[(i)]
\begin{align}\label{15.3}
V_{f\wedge g}(i)=V_f(i)\wedge V_g(i),\quad i=1,2,\cdots,2^n.
\end{align}
\item[(ii)]
\begin{align}\label{15.4}
V_{f\vee g}(i)=V_f(i)\vee V_g(i),\quad i=1,2,\cdots,2^n.
\end{align}
\item[(iii)]
\begin{align}\label{15.5}
V_{\neg f}(i)=\neg f(i), \quad i=1,2,\cdots,2^n.
\end{align}
\item[(iv)]
\begin{align}\label{15.6}
V_{\bf 1}={\bf 1}_{2^n}^\mathrm{T},\quad V_{\bf 0}={\bf 0}_{2^n}^\mathrm{T}.
\end{align}
\end{itemize}
\end{dfn}

The following result comes from Definition \ref{d.15.1} immediately:

\begin{prp}\label{p.15.4} Consider BN  (\ref{2.1.1}) with its ASSR (\ref{2.1.4}).
\begin{itemize}
\item[(i)] ${\cal X}^*$ with $\wedge$, $\vee$, $\neg$, ${\bf 1}$, and ${\bf 0}$, defined in Definition \ref{d.15.2}, is a Boolean algebra.
\item[(ii)] Let $M\in {\cal L}_{2^n\times 2^n}$. Then $\psi_M:{\cal X}^*\ra {\cal X}^*$, defined as,
\begin{align}\label{15.7}
V_{\psi_M(f)}=V_fM,
\end{align}
is a ${\cal B}_A$ homomorphism.
That is,
\begin{align}\label{15.8}
\begin{array}{l}
\psi_M({\bf 1})={\bf 1},\\
\psi_M({\bf 0})={\bf 0},\\
\psi_M(x^*\wedge y^*)=\psi_M(x^*)\wedge \psi_M(y^*),\\
\psi_M(x^*\vee y^*)=\psi_M(x^*)\vee \psi_M(y^*),\\
\psi_M(\neg x^*)=\neg \psi_M(x^*).\\
\end{array}
\end{align}
\end{itemize}
\end{prp}

\begin{dfn} \label{d.15.5} Let
$$
d^*_i(x)=
\begin{cases}
1,\quad x=\d_{2^n}^i\\
0,\quad \mbox{Otherwise}
\end{cases}
\in {\cal X}^*,\quad i=1,2,\cdots,2^n.
$$
Then $\{d_i^*\;|\;i=1,2,\cdots,2^n\}$
is called a set of generators of ${\cal X}^*$.
\end{dfn}

\begin{prp}\label{p.15.6}
Let
$$V_f=\bigvee_{j=1}^s V_{d^*_{i_j}}.
$$
Then
\begin{align}\label{15.9}
f(x)=\bigvee_{j=1}^s d^*_{i_j}(x).
\end{align}
Moreover,
if $f(t)=\bigvee_{j=1}^s d^*_{i_j}$, then
\begin{align}\label{15.10}
f(t+1)=\bigvee_{j=1}^s V_{d^*_{i_j}}M.
\end{align}
\end{prp}

(\ref{15.10}) can be used to construct the dynamics on ${\cal X}^*$.

\begin{exa}\label{e.15.7} Recall Example \ref{e4.4}.
It is easy to calculate that
$$
\begin{array}{l}
V_{d_1^*}M=[1,0,0,0,0,0,0,0]M=[0,1,0,0,0,0,0,0],\\
V_{d_2^*}M=[0,1,0,0,0,0,0,0]M=[1,0,1,0,0,0,0,0],\\
V_{d_3^*}M=[0,0,1,0,0,0,0,0]M=[0,0,0,0,0,0,0,0],\\
V_{d_4^*}M=[0,0,0,1,0,0,0,0]M=[0,0,0,1,1,0,0,0],\\
V_{d_5^*}M=[0,0,0,0,1,0,0,0]M=[0,0,0,0,0,0,0,0],\\
V_{d_6^*}M=[0,0,0,0,0,1,0,0]M=[0,0,0,0,0,0,1,0],\\
V_{d_7^*}M=[0,0,0,0,0,0,1,0]M=[0,0,0,0,0,0,0,1],\\
V_{d_8^*}M=[0,0,0,0,0,0,0,1]M=[0,0,0,0,0,0,0,0],\\
\end{array}
$$
Now assume
$$
V_{f(t)}=[1,0,0,0,1,0,1,0]=V_{d_1^*}\bigvee V_{d_5^*}\bigvee V_{d_7^*}.
$$
Then
$$
\begin{array}{ccl}
V_{f(t+1)}&=&V_{d_1^*}M\bigvee V_{d_5^*}M\bigvee V_{d_7^*}M\\
~&=&[0,1,0,0,1,0,0,1].
\end{array}
$$

Using the vector form provided in equation (\ref{3.1.0.1}), we have
$$
\begin{array}{l}
V_{f(t)}=V_{x^*_{139}},\\
V_{f(t+1)}=V_{x^*_{74}},
\end{array}
$$
hence
$$
x^*_{139}(t+1)=x^*_{74}(t).
$$
Similarly, all the dynamic equations of ${\cal X}_i^*$ can be obtained.

\end{exa}

\begin{rem}\label{r.15.8}According to this Boolean algebra structure, among $2^{2^n}$ logical equations in ${\cal X}^*$, only $2^n$ are independent, which compose a coordinate transformation of ${\cal X}$.
\end{rem}
\section{Realization of BCN}

Consider BCN
\begin{align}\label{6.1}
\begin{array}{l}
\begin{cases}
X_1(t+1)=f_1(X_1(t),\cdots,X_n(t),U_1(t),\cdots,U_m(t))\\
X_2(t+1)=f_2(X_1(t),\cdots,X_n(t),U_1(t),\cdots,U_m(t))\\
\vdots\\
X_n(t+1)=f_n(X_1(t),\cdots,X_n(t),U_1(t),\cdots,U_m(t)),\\
\end{cases}\\
~~~Y_j(t)=g_j(X_1(t),\cdots,X_n(t)),\quad j\in [1,p],
\end{array}
\end{align}
where $X_i(t)$, $i\in[1,n]$ are state variables, $U_k(t)$, $k\in[1,m]$ are controls, $Y_j(t)$, $j\in[1,p]$ are outputs.
Its  ASSR is described as
\begin{align}\label{6.2}
\begin{array}{l}
x(t+1)=Lu(t)x(t),\\
y(t)=Ex(t),
\end{array}
\end{align}
where $x(t)=\ltimes_{i=1}^nx_i(t)$, $u(t)=\ltimes_{k=1}^mu_k(t)$, $y(t)=\ltimes_{j=1}^py_j(t)$,
$L\in {\cal L}_{2^n\times 2^{m+n}}$, $E\in {\cal L}_{2^p\times 2^{n}}$.

\begin{dfn}\label{d.6.1} Consider BCN (\ref{6.1}) with its ASSR (\ref{6.2}).

 Let $\Xi^*\subset {\cal X}^*$. If ${\cal V}^*\subset {\cal X}^*$ satisfies
\begin{itemize}
\item[(i)]
$
\Xi^*\subset {\cal V}^*.
$
\item[(ii)]
${\cal V}^*$ is $M_i$-invariant, where $M_i=L\d_{2^m}^i$, $i\in[1,2^m]$,
\end{itemize}
then ${\cal V}^*$ is a control invariant subspace (CIS) containing $\Xi^*$.

If ${\cal V}^*$ is a CIS, and
\begin{itemize}
\item[(iii)]
for any other CIS ${\cal W}^*$, ${\cal V}^*\subset {\cal W}^*$,
\end{itemize}
then ${\cal V}^*$ is called the smallest CIS containing $\Xi^*$.
\end{dfn}
Now we provide an algorithm to calculate the smallest CIS containing a given $\Xi^*$.

\begin{algorithm}
\KwData{$(\Xi^*,L,m,n)$
\\\%a subset of the dual space ${\cal X}^*$, the structure matrix \\\%of the BCN, the number of controls, the number of \\\%state variables}
 \KwResult{${\cal V}^*$\\\%the smallest CIS containing $\Xi^*$}
 \BlankLine
 $j=0,k=0,{\cal V}_0^*:=\Xi^*$\;
 \While{$k < 2^{2^n}$}
{${\cal V}^*_{k+1}:={\cal V}^*_k\bigcup\left[\cup_{j=1}^{2^m}{\cal V}_k^*L\delta_{2^m}^j\right]$\;
  \eIf{${\cal V}^*_{k+1}={\cal V}^*_{k}$}{${\cal V}^*:={\cal V}^*_k$\;$k:=2^{2^n}$\;}{$k++$;}
 }
 \caption{Smallest CIS containing $\Xi^*$\label{a.6.2}}
\end{algorithm}

From Definition \ref{d.6.1} and Algorithm \ref{a.6.2}, the following conclusion can be easily obtained.
\begin{prp}\label{p.6.3}
${\cal V}^*$ obtained by Algorithm \ref{a.6.2} is the CIS containing $\Xi^*$.
\end{prp}

Assume ${\cal V}^*=\{z_1,z_2,\cdots,z_s\}$ and
$$
z_i=G_ix,\quad i=1,2,\cdots,s.
$$
Let $z=\ltimes_{i=1}^sz_i$. Then
$$
z=Gx,
$$
where $G=G_1*G_2*\cdots*G_s\in {\cal L}_{2^s\times 2^n}$.

Next, consider
\begin{align}\label{6.11}
\begin{array}{ccl}
z(t+1)&=&Gx(t+1)\\
~&=&GLu(t)x(t)\\
~&=&\left[GM_1,GM_2,\cdots,GM_{2^m}\right]u(t)x(t),
\end{array}
\end{align}

Since ${\cal V}^*$ is $M_i$-invariant, there exists $H_i$, such that
$$
GM_i=H_iG,\quad i=1,2,\cdots,2^m.
$$
Set
$$
H=[H_1,H_2,\cdots,H_{2^m}]
$$
Then (\ref{6.11}) becomes
\begin{align}\label{6.12}
z(t+1)=Hu(t)z(t).
\end{align}

Consider BCN (\ref{6.1}) with its ASSR (\ref{6.2}). Now we suppose that $\rho_i\in {\cal V}^*$, $i\in[1,p]$. Then the outputs can be expressed by
$$
y_i=\rho_i(x)=F_iz,\quad i=1,2,\cdots,p.
$$
Hence,
\begin{align}\label{6.13}
y(t)=Fz(t),
\end{align}
where $F=F_1*F_2*\cdots*F_p$.

Summarizing the above, we have the following concept as a matter of course.

\begin{dfn}\label{d.6.4} Let ${\cal V}^*\subset {\cal X}^*$ be the smallest CIS, which contains $y_j=\rho_j(x)$, $j=1,2,\cdots,p$ and is $M_i$-invariant, $i=1,2,\cdots,2^m$. Then the corresponding system (\ref{6.12})-(\ref{6.13}) is called the minimum realization of BCN (\ref{6.1}).
\end{dfn}

\begin{rem}\label{r.6.5}
\begin{itemize}
\item[(i)] The minimum realization is on ${\cal X}^*$. Unlike continuous (control) systems, since
$$
\left|{\cal X}^*\right|>> \left|{\cal X}\right|,
$$
the dimension of minimum realization may be larger than the dimension of the original BCN.
\item[(ii)] Using the Boolean algebraic structure, the dimension of minimum realization can further be reduced, which assumes the dimension of the minimum realization being less than or equal to the dimension of the original BCN.
\end{itemize}
\end{rem}

\begin{exa}\label{e.6.6} Consider the following BCN
\begin{align}\label{6.14}
\begin{cases}
x(t+1)=Lu(t)x(t),\\
y(t)=Ex(t),
\end{cases}
\end{align}
where $x(t)\in \D_8$, $u(t),~y(t) \in \D_2$,
$$
L=\d_8[4,2,8,8,5,6,6,3,8,7,3,1,4,6,6,4],
$$
$$
E=\d_2[2,2,2,2,1,2,2,1].
$$

 Using Algorithm \ref{a.6.2}, we have
the smallest CIS containing $y$ as
$$
{\cal V}^*=\{x_1^*=y,x_2^*,x_3^*\},
$$
where
$$
\begin{array}{l}
x_2^*=\d_2[2,1,2,2,2,2,2,2]x,\\
x_3^*=\d_2[2,2,2,2,2,2,2,2]x.
\end{array}
$$

The minimum realization of (\ref{6.14}) is
\begin{align}\label{6.15}
\begin{array}{l}
\begin{cases}
x^*_1(t+1)=x^*_2(t),\\
x^*_2(t+1)=[x_2^*(t),x^*_3(t)]u(t),\\
x^*_3(t+1)=\d_2^2.
\end{cases}\\
~~~y(t)=x^*_1(t).
\end{array}
\end{align}
\end{exa}

\begin{figure}
\centering
\setlength{\unitlength}{0.49 cm}
\begin{picture}(18,15)\thicklines
\put(8,1){\line(1,0){4}}
\put(4,3){\line(1,0){8}}
\put(4,5){\line(1,0){12}}
\put(4,7){\line(1,0){12}}
\put(4,9){\line(1,0){12}}
\put(6,11){\line(1,0){8}}
\put(8,13){\line(1,0){4}}
\put(4,5){\line(0,1){4}}
\put(6,3){\line(0,1){8}}
\put(8,1){\line(0,1){12}}
\put(10,1){\line(0,1){12}}
\put(12,1){\line(0,1){12}}
\put(14,3){\line(0,1){8}}
\put(16,5){\line(0,1){4}}
\put(2,9){\vector(1,0){2}}
\put(4,7){\vector(-1,0){2}}
\put(16,11){\vector(-1,0){2}}
\put(12,13){\vector(1,0){2}}
\put(14,1){\vector(-1,0){2}}
\put(14,3){\vector(1,0){2}}
\put(2.2,9.2){$u^1$}
\put(2.2,7.4){$y^1$}
\put(15.6,11.2){$u^2$}
\put(13.6,13.4){$y^2$}
\put(13.6,1.2){$u^s$}
\put(15.6,3.4){$y^s$}
\thinlines
\put(1,4){\line(1,0){4}}
\put(1,4){\line(0,1){6}}
\put(5,10){\line(-1,0){4}}
\put(5,10){\line(0,-1){6}}
\put(11,10){\line(1,0){6}}
\put(11,10){\line(0,1){4}}
\put(17,14){\line(-1,0){6}}
\put(17,14){\line(0,-1){4}}
\put(13,0){\line(1,0){4}}
\put(13,0){\line(0,1){4}}
\put(17,4){\line(-1,0){4}}
\put(17,4){\line(0,-1){4}}
\thicklines
\put(1.2,4.2){$\Sigma_1$}
\put(16,13.2){$\Sigma_2$}
\put(16,0.2){$\Sigma_s$}
\end{picture}
\caption{Distributed Realization\label{Fig.6.1}}
\end{figure}
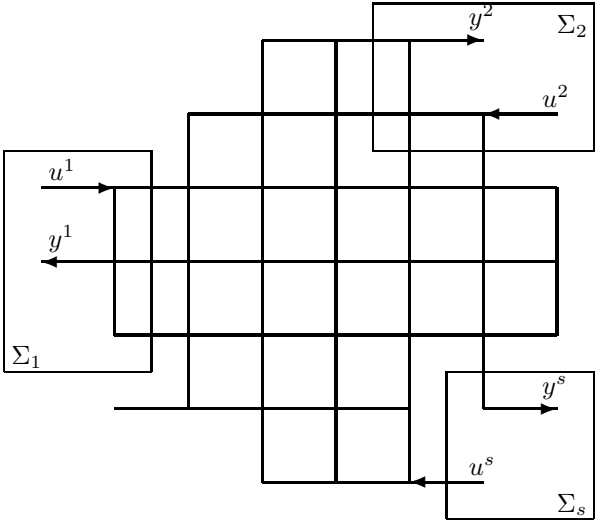

Consider a large-scale BN (please refer to Fig. \ref{Fig.6.1}).

A distributed realization is described as follows:

\begin{itemize}
\item[(i)] Inject some inputs:
$$
u^i,\quad i=1,2,\cdots,s.
$$
\item[(ii)]
Observe some dates:
$$
y^j,\quad i=1,2,\cdots,s.
$$
\item[(iii)]
Consider the corresponding minimum realizations
$$
\Sigma_i:u^i\Rightarrow y^i,\quad i=1,2,\cdots,s.
$$
\end{itemize}

We can investigated parts of the BN for some particular properties.
This kind of realizations is called the distributed realization, which is helpful for investigating large-scale BNs.

\section{$k$-Valued Logical Network}

Denote
$$
{\cal D}_k:=\{0,\frac{1}{k-1},\frac{2}{k-1},\cdots,1\}, \quad k\geq 2.
$$
When $k=2$, ${\cal D}_2={\cal D}$. This subsection consider the case when $k\geq 3$.

Setting
$$
\frac{i}{k-1}\sim \d_{k}^{k-i},\quad i=0,1,\cdots,k-1,
$$
we have the vector expression of $X\in {\cal D}_k$ as $x=\vec{X}\in \D_k$.

Equation (\ref{2.1.1}) is a $k$-valued logical network, if $X_i\in {\cal D}_k$ and $f_i:{\cal D}_k^n\ra {\cal D}_k$, $i=1,2,\cdots,n$. Using vector form expression, we also have (\ref{2.1.2}) as its ASSR, where $M\in {\cal L}_{k^n\times k^n}$.

The state space is
$$
{\cal X}=\{(X_1,X_2,\cdots,X_n)\;|\;X_i\in {\cal D}_k,\; i=1,2,\cdots,n\}\sim \D_{k^n}.
$$

Let $f:{\cal D}_k^n\ra {\cal D}_k$. Similarly to Boolean case, there exists a unique logical matrix
$M_f\in {\cal L}_{k\times k^n}$ such that
$$
f(x)=M_fx=\d_{k}[i_1,i_2,\cdots,i_{k^n}]x.
$$
The vector
$$
V_f:=[i_1,i_2,\cdots,i_{k^n}]
$$
is called the structure vector of $f$.

Its dual space is
$$
\begin{array}{ccl}
{\cal X}^*&=&\{f\;|\; f:{\cal D}_k^n\rightarrow {\cal D}_k\}\\
&\sim&\left\{[i_1,i_2,\cdots,i_{k^n}]\;|\; 1\leq i_j\leq k,\;j\in[1,k^n]\right\}.
\end{array}
$$

An argument similar to Boolean case, one sees that aforementioned arguments about Boolean (control) networks
with certain obvious modification remain true for $k$-valued logical networks.

\section{Concluding Remarks}

By introducing the dual space, dual BN, and the minimum realization of BCNs, the hidden order of a BN is firstly revealed and explored.

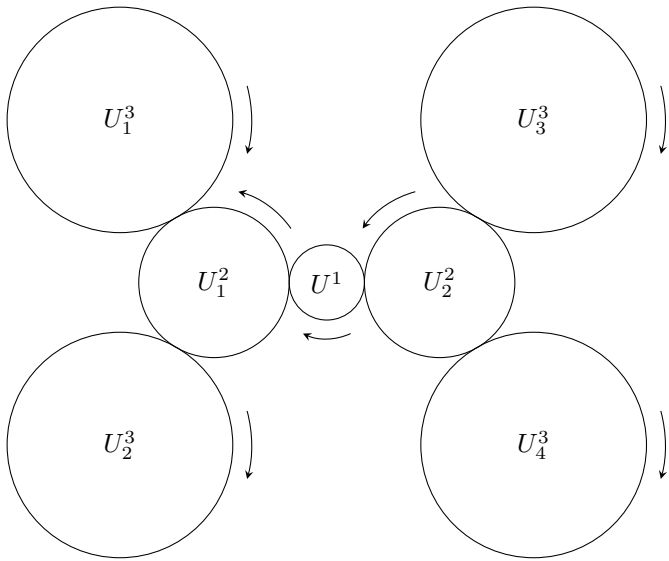
\begin{figure}[!ht]
\centering
\begin{tikzpicture}[>=stealth,zoom/.style={scale=.75}]
\coordinate (O) at (0,0);
\coordinate (OL) at (-1.5,0);
\coordinate (OR) at (1.5,0);
\coordinate (OLT) at (-2.75,2.1625);
\coordinate (OLB) at (-2.75,-2.1625);
\coordinate (ORT) at (2.75,2.1625);
\coordinate (ORB) at (2.75,-2.1625);
\node at (O) {$U^{1}$};
\node at (OL) {$U_{1}^{2}$};
\node at (OR) {$U_{2}^{2}$};
\node at (OLT) {$U_{1}^{3}$};
\node at (OLB) {$U_{2}^{3}$};
\node at (ORT) {$U_{3}^{3}$};
\node at (ORB) {$U_{4}^{3}$};
\draw (O) circle (.5cm); \draw [<-] (O)+(245:.75cm) arc (245:295:.75cm);
\draw (OL) circle (1cm); \draw [->] (OL)+(35:1.25cm) arc (35:75:1.25cm);
\draw (OR) circle (1cm); \draw [->] (OR)+(105:1.25cm) arc (105:145:1.25cm);
\draw (OLT) circle (1.5cm); \draw [->] (OLT)+(15:1.75cm) arc (15:-15:1.75cm);
\draw (OLB) circle (1.5cm); \draw [->] (OLB)+(15:1.75cm) arc (15:-15:1.75cm);
\draw (ORT) circle (1.5cm); \draw [->] (ORT)+(15:1.75cm) arc (15:-15:1.75cm);
\draw (ORB) circle (1.5cm); \draw [->] (ORB)+(15:1.75cm) arc (15:-15:1.75cm);
\end{tikzpicture}
\caption{Chained Gears Structure of Attractors\label{Fig.9.1}}
\end{figure}

It was pointed out by Kauffman that \cite{kau95} the tiny attractors in a large scale Boolean network determine the vast order. In \cite{che09} the structure of chained gears was proposed to explain why tiny attractors determine the order of overall BN. Fig. \ref{Fig.9.1} depicts a set of chained gears, where each circle represents a cycle. 

In the structure of chained gears, the tiny gears can be considered as the driving gears, and the large gears can be considered as the following ones. 
Hence the tiny gears determine the order of overall system.

In fact, this structure appears only to dual space ${\cal X}^*$. Observing Example \ref{e4.4} again. Assume $z_i\in {\cal X}^*_i$, $i=1,2,3$ are lying on (dual) cycles $C^*_i\subset {\cal X}_i^*$ respectively, and $|C^*_i|=\ell_i$. If $z_{1+2}=z_1\vee z_2$, then it is obvious that $z_{1+2}$ is on a cycle $C^*_{1+2}\in {\cal X}^*_1\cup {\cal X}^*_2$, and $|C^*_{1+2}|=\lcm(\ell_1,\ell_2)$. Furthermore, If $z_{1+2+3}=z_1\vee z_2\vee z_3$, then $C^*_{1+2+3}\in {\cal X}^*_1\cup {\cal X}^*_2\cup {\cal X}^3$ can be obtained. In general, a large scale cycle in ${\cal X}^*$ can be generated by tiny cycles within each invariant (dual) subspaces ${cal X}^*_i$. It leads to the conclusion that the tiny attractors determine the vast order. Moreover, this fact also reveals that hidden order from dual NB may play more important role for determining the order of BNs, while it is used to model live world. 

It seems to us that the (explicit) order determined by the attracts of a BN is the inner order of a BN. The order observed by us may be the order of observed functions, which comes exactly from the dual BN.  

A DNA system with $A,T,G,C$ may be considered as a $4$ valued network. It is our conjecture that the hidden order from its dual network might be the key to understand it.


\begin{thebibliography}{00}

%
%
\bibitem{cas19} F.Z. Castro, M.E.Valle, A broad class of discrete-time hypercomplex-valued Hopfield neuralnetworks, {\it arXiv: 1902.05478v3 [cs.LG] 31 Oct, 2019}, 2019.
%
\bibitem{che09} D. Cheng, Input-state approach to Boolean networks, {\it IEEE Trans.  Neural Networks}, Vol. 20, No. 3, 512-521,  2009.
%
\bibitem{che10} D. Cheng, H. Qi, State-space analysis of Boolean networks, {\it IEEE Trans. Neural Networks}, Vol. 21, No. 4, 584-594,  2010.
%
\bibitem{che11} D. Cheng, H. Qi, Z. Li, {\it Analysis and Control of Boolean Networks - A Semi-tensor Product Approach}, Springer, London, 2011.
%
\bibitem{che12} D. Cheng, H. Qi, Y. Zhao, {\it An Introduction to Semi-tensor Product of Matrices and Its Applications}, World Scientific, Singapore, 2012.
%
%
\bibitem{chepr} D. Cheng, L. Zhang, D. Bi, Invariant subspace approach to Boolean (Control) Networks, {\it IEEE Trans. Aut. Contr.}, (provisionary accepted).

\bibitem{for13} E. Fornasini, M.E. Valcher, Observability, reconstructibility and state observers of Boolean control networks, {\it IEEE Trans. Aut. Contr.}, Vol. 58, No. 6, 1390-1401, 2013.

\bibitem{hol95} J.H. Holland, {\it Hidden Order}, Addison-Wesley Pub. Comp., New York, 1995.

\bibitem{hor85} R.A. Horn, C.R. Johnson, {\it Matrix Analysis}, Cambridge Univ. Cambridge, England, 1985.
\bibitem{kau69} S.A. Kauffman, Matabolic stability and epigenesis in randomely connected nets, {\it J. Theoret. Biol.}, Vol. 22, 437, 1969.

\bibitem{kau93} S.A. Kauffman, {\it The Origins of Order}, Oxford Univ. Press, New York, 1993.
%
\bibitem{kau95} S.A. Kauffman, {\it At Home in the Universe}, Oxford Univ. Press, New York, 1995.
%
\bibitem{liu19} Z. Liu, D. Cheng, Canonical form of Boolean networks, {\it Proc. 2019 CCC}, 1801-1806, 2019.

\bibitem{ros03} K.A. Ross, C.R.B. Wright, {\it Discrete Mathematics}, 5th Ed.,  Prentice Hall, New York, 2003.
%

\bibitem{wal95} S.A. Kauffman, {\it At Home in the Universe: The Search for the Laws of Self-Organization and Complexity}, Oxford University Press, New York, 1995.



\end{thebibliography}
\end{document}